\documentclass[12pt]{article}
 
\usepackage{amsmath,graphicx}
\usepackage{multirow}
\usepackage{amsfonts}
\usepackage{amssymb}
\usepackage{amscd}
\usepackage{cite}
\usepackage{amsmath}
\usepackage{bbm}
\usepackage{color}
\usepackage{comment}
\usepackage{hyperref}
\hypersetup{
    colorlinks,
    citecolor=black,
    filecolor=black,
    linkcolor=black,
    urlcolor=black
}

\def\hybrid{\topmargin -20pt    \oddsidemargin 0pt
        \headheight 0pt \headsep 0pt
        \textwidth 6.25in       
        \textheight 9.5in       
        \marginparwidth .875in
        \parskip 5pt plus 1pt   \jot = 1.5ex}

\hybrid

\numberwithin{equation}{section}
\numberwithin{table}{section}

\renewcommand{\Re}{\operatorname{Re}}
\renewcommand{\Im}{\operatorname{Im}}
\newcommand\e{\mathrm{e}}

\newcommand{\M}{{\bf M}}
\def\Mq{\hat{\M}}
\newcommand\diff{\mathrm{d}}
\newcommand\iu{\operatorname{i}}
\def\ax{{\tilde \phi}} 
\def\Proj{{\Pi}}
\def\cN{{\cal N}}
\def\cF{{\cal F}}
\def\cG{{\cal G}}
\def\nv{n_{\rm v}}
\def\nh{n_{\rm h}}
\def\nqv{{\hat n}_{\rm v}}
\def\nqh{{\hat n}_{\rm h}}
\def\mino{m_{3/2}}
\def\cN{{\cal N}}
\def\ax{{\tilde \phi}} 
\def\Kdil{{k_\phi}}    
\def\Kax{{k_{\tilde \phi}}}  
\def\Kxi{{k}}                
\def\Ktxi{{\tilde k}}        
\def\kk{{k}}          
\def\kkk{{k}}          

\def\hatP{{U}}  
\def\hatQ{{V}}

\begin{document}
\begin{titlepage}
\begin{center}
\rightline{\small ZMP-HH/13-9}
\vskip 1cm

{\Large \bf
Examples of ${\cal N} = 2$ to ${\cal N} = 1$ supersymmetry breaking}	
\vskip 1.2cm
{\bf  Tobias Hansen$^{a}$ and Jan Louis$^{a,b}$ }

\vskip 0.8cm

$^{a}${\em II. Institut f\"ur Theoretische Physik der Universit\"at Hamburg, Luruper Chaussee 149, D-22761 Hamburg, Germany}
\vskip 0.4cm
$^{b}${\em Zentrum f\"ur Mathematische Physik,
Universit\"at Hamburg,\\
Bundesstrasse 55, D-20146 Hamburg, Germany}
\vskip 0.8cm

{\tt tobias.hansen@desy.de, jan.louis@desy.de}

\end{center}

\vskip 1cm
\vskip 1cm

\begin{center} {\bf ABSTRACT } \end{center}

\noindent

In this paper we consider gauged  $\cN=2$ supergravities which arise in the
low-energy limit of type II string theories and study
examples
which exhibit spontaneous partial supersymmetry breaking.
For the  quantum $STU$ model we derive the  scalar field space and the scalar
potential of the 
$\cN=1$ supersymmetric low-energy effective action.
We also  study the properties of the Minkowskian $\cN=1$ supersymmetric
ground states 
for a broader class of supergravities including the 
 quantum $STU$ model.
\vfill

June 2013

\end{titlepage}

\tableofcontents
\newpage

\section{Introduction}
\label{sec:intro}

After the initial no-go theorem of Refs.~\cite{Cecotti:1984rk,Cecotti:1984wn}
explicit $\cN=2$ supergravities with spontaneous partial supersymmetry breaking 
were first discussed in
Refs.~\cite{Ferrara:1995gu,Ferrara:1995xi,Fre:1996js} 
following the global analysis of
Refs.~\cite{Bagger:1994vj,Antoniadis:1995vb}.
Recently a systematic analysis was performed 
in Refs.~\cite{LST1,LST2,CLST} using the embedding tensor formalism
\cite{deWit:2005ub}.
It was shown that  both electrically and magnetically charged
hypermultiplets have to be in the spectrum in order to circumvent the 
no-go theorem of Refs.~\cite{Cecotti:1984rk,Cecotti:1984wn}.
The presence of two holomorphic commuting Abelian isometries in the
hypermultiplet sector
are an additional necessary ingredient of the supergravity to show partial
supersymmetry breaking.

Below the scale of the supersymmetry breaking $m_{3/2}$
one can construct
a low-energy effective ${\cal N} = 1$ action 
in terms of the $\cN=2$ ``input data''  by integrating out
all heavy modes with masses of order ${\cal O}(m_{3/2})$.
The resulting ${\cal N} = 1$ field space is a quotient of the original
${\cal N} = 2$ quaternionic-K\"ahler manifold with respect to the
two isometries.
The properties of the two isometries ensure that the quotient
is a K\"ahler manifold 
consistent with the  ${\cal N} = 1$ of the effective theory \cite{LST2,CLST}.
If only these two isometries are gauged, no further supersymmetry breaking 
is possible.
However, if additional  isometries, which do not participate in the  
$\cN=2\to\cN=1$ breaking, are gauged, the superpotential ${\cal W}$ 
and the $\cal D$-terms
can be non-trivial and possibly induce the breaking of $\cN=1$ 
at a lower scale.

In this paper we consider  $\cN=2$ supergravities which arise as the 
low energy effective theory of type II string
compactifications, i.e., supergravities which are in the image
of the c-map \cite{Cecotti:1988qn}. 
In this class of  supergravities the
 quaternionic-K\"ahler manifold of the hypermultiplets 
is `special' in that it has a specific fibration structure
with a base which is
determined by a holomorphic prepotential ${\cal G}$.
As a consequence of the fibration
isometries exist which can induce partial supersymmetry breaking. 

%
In the resulting ${\cal N}=1$ backgrounds some of the original
scalar fields of the ${\cal N}=2$ theory acquire a vacuum expectation value
and become massive.
For special quaternionic-K\"ahler manifolds only two scalars of the fibre
are fixed while in the base the number of massive scalars depends 
on the form of the prepotential ${\cal G}$.
For a quadratic ${\cal G}$ no scalars are fixed while
a generic ${\cal G}$ fixes all scalars in the base. 
We focus on 
two specific examples where the base of the fibration
is a complex three-dimensional manifold determined 
by the cubic $STU$ and quantum
$STU$ prepotentials. In both models one complex scalar field of the base 
together with three fields in the fibre
remain free and we explicitly determine ${\cal W}$ and ${\cal D}$
below the scale of partial supersymmetry breaking together with the 
${\cal N}=1$ scalar field space.
For the $STU$-model we find the latter to be the symmetric space
 $\frac{SO(4,2)}{SO(4) \times SO(2)}\,$.

We also discuss 
the conditions for ${\cal N}=1$ supersymmetric Minkowskian vacua 
for generic supergravities in the image of the c-map.
The superpotential and the $\cal D$-term 
depend on two holomorphic prepotentials, 
the ${\cal G}$ of the hypermultiplet sector
together with the 
${\cal F}$ which encodes the couplings of the vector multiplets.
We show that the form of these prepotentials does not only determine the dimension 
of the ${\cal N}=1$ scalar field space but also the existence
of ${\cal N}=1$ supersymmetric minima.

This paper is organized as follows. 
In Section \ref{sec:recapSection} we recapitulate the relevant aspects
of four-dimensional gauged ${\cal N} = 2$ supergravities,
partial  ${\cal N} = 2\to{\cal N} = 1$ supersymmetry breaking and the resulting ${\cal N} = 1$ low energy effective action.
In Section \ref{sec:specialQuaternionicKaehler} we focus on special quaternionic-K\"ahler manifolds which arise in the hypermultiplet sector of  type II string compactifications and recall for this case the  quotient construction
leading to the $\cN=1$ low energy effective action.
In Section \ref{sec:STU} we calculate  the $\cN=1$
K\"ahler potential explicitly
for 
the (quantum) $STU$ model.
In Section~\ref{WD} we compute the superpotential and the $\cal D$-terms
for  generic special quaternionic-K\"ahler manifolds and for the 
quantum $STU$-model. In
Section \ref{sec:generalVacuum} we determine the supersymmetric ground states
of the ${\cal N} = 1$ theory.
Appendix \ref{sec:field_redefinitions} discusses field redefinitions that are used to bring the quantum $STU$ K\"ahler potential to its final form and
Appendix \ref{sec:the_killing_prepotentials} contains the calculation of Killing prepotentials that appear in the superpotential and $\cal D$-terms.

\section{Partially broken \texorpdfstring{${\cal N} = 2$}{N = 2} supergravities}
\label{sec:recapSection}

\subsection{\texorpdfstring{${\cal N} = 2$}{N = 2} supergravity in four dimensions}

In order to set the stage let us briefly recall
some properties of four-dimensional ${\cal N} = 2$ supergravity with 
gauged Abelian isometries in the hypermultiplet sector
(for a review, see e.g.\ \cite{Andrianopoli:1996cm}).
The spectrum consists of a gravitational multiplet, $\nv$ vector
multiplets and $\nh$ hypermultiplets. The gravitational multiplet
$(g_{\mu\nu},\Psi_{\mu {\cal A}}, A_\mu^0)$ contains the spacetime
metric $ g_{\mu\nu}, \mu,\nu =0,\ldots,3$, two gravitini $\Psi_{\mu
  {\cal A}}, {\cal A}=1,2$, and the graviphoton $A_\mu^0$. A~vector
multiplet $(A_\mu,\lambda^{\cal A}, t)$ contains a vector $A_\mu$, two
gaugini $\lambda^{\cal A}$  and a complex scalar $t$, while a
hypermultiplet $(\zeta_{\alpha}, q^u)$ contains two hyperini
$\zeta_{\alpha}$ and four real scalars $q^u$. 
%

The Lagrangian of the scalar fields
is given by
 \begin{equation}\begin{aligned}\label{sigmaint}
{\cal L}\ =\  
 g_{i\bar j}(t,\bar t)\, \partial_\mu t^i \partial^\mu\bar t^{\bar j}
+ h_{uv}(q)\, D_\mu q^u D^\mu q^v
- V(t,q) 
\ ,
\end{aligned}\end{equation}
where the indices take the values $i,\bar j=1,\ldots,\nv,\, u,v=1,\ldots, 4 \nh$.
$g_{i\bar j}(t,\bar t)$  is the metric of a $2\nv$-dimensional special-K\"ahler manifold ${\M}_{\rm v}$ 
and $h_{uv}(q)$ is the metric of the $4\nh$-dimensional quaternionic-K\"ahler manifold ${\M}_{\rm h}$.
From \eqref{sigmaint} we see that the total scalar field space locally is the direct product
\begin{equation}\label{scalarM}
{\M} = {\M}_{\rm v}\times {\M}_{\rm h} \ .
\end{equation}

The metric of the special-K\"ahler manifold is given by 
\begin{equation}\label{gdef}
g_{i\bar j} = \partial_i \partial_{\bar j} K^{\rm v}\ ,
\quad \textrm{where }\quad
K^{\rm v}= -\ln \iu \left( \bar X^I \cF_I - X^I\bar \cF_I \right)\ 
\end{equation}
is the K\"ahler potential.
$X^I(t)$ and ${\cal F}_I(t)$, $I= 1,\ldots,\nv+1,$ are holomorphic 
functions of the~$t^i$
 with $\cF_I = \partial\cF/\partial{X^I}$ being the derivatives of a
 holomorphic prepotential ${\cal F}(X)$ which is homogenous of degree
 two. 
There is a choice of coordinates, called ``special coordinates'', where 
$X^I = (t^i,1)$.\footnote{We are choosing this convention in order to be consistent with Ref.~\cite{CLST} later on.}  


$h_{uv}$ in \eqref{sigmaint} denotes the metric on the
$4\nh$-dimensional quaternionic-K\"ahler manifold ${\M}_{\rm h}$.
These manifolds admit three almost complex structures $J^x,\, x=1,2,3$ that satisfy the quaternionic algebra
\begin{equation}\label{jrel}
J^x J^y = -\delta^{xy}{\bf 1} + \epsilon^{xyz} J^z ~,
\end{equation}
with the metric $h_{uv}$ being hermitian with respect to all
three of them.
Supersymmetry requires the existence of a principal $SU(2)$-bundle 
over ${\M}_{\rm h}$
with a  curvature two-form 
\begin{equation} \label{def_Sp(1)_curvature}
 K^x =\diff \omega^x + \tfrac12 \epsilon^{xyz} \omega^y\wedge \omega^z\ ,
\end{equation}
where $\omega^x$
denotes the one-form connection on the $SU(2)$-bundle.


For simplicity we consider only Abelian gauge fields and hence their supersymmetric scalar partners $t^i$ are neutral. The scalars $q^u$ in the hypermultiplets
on the other hand can be charged and  their covariant derivatives 
in \eqref{sigmaint} are defined as
\begin{equation}\label{d2}
D_{\mu} q^u
= \partial_{\mu}  q^u - A^{~I}_{\mu}\, \Theta_I^{~\lambda}\, \kk_{\lambda}^u +
B_{\mu I}\, \Theta^{I{\lambda}}\, \kk_{\lambda}^u \ ,
\end{equation}
where $\lambda$ labels the
non-trivial Killing vectors $\kk_\lambda(q)$ on $\M_{\rm h}$ and 
$A_\mu^{I}$ are electric vectors while 
$B_{\mu I}$ are their magnetic duals. The charges
$\Theta_I^{~\lambda}$ and $\Theta^{I{\lambda}}$ are the electric and
magnetic parts of the embedding tensor which in the following 
we frequently combine into the
symplectic object $\Theta_\Lambda^{~\lambda} = (\Theta_I^{~\lambda},
-\Theta^{I{\lambda}})$~\cite{deWit:2005ub}. 
Mutual locality of electric and magnetic
charges imposes
\begin{equation}
\Theta^{I[{\lambda}} \Theta_{I}^{\phantom{I}\kappa]} = 0\;.
\label{embeddingTensorLocality}
\end{equation}

Finally,
the scalar potential $V$ appearing in \eqref{sigmaint} is given by 
\begin{equation} \label{potential_identity}
 V = -12 S_{\cal AB} \bar S^{\cal AB} + g_{i\bar \jmath} W^{i \cal {AB}} W^{\bar \jmath}_{\cal AB}
+ 2 N_\alpha^{\cal A} N_{\cal A}^\alpha \ ,
\end{equation}
where the couplings $S_{\cal AB}, W^{i{\cal AB}}$ and $N_\alpha^{\cal A}$ 
denote the scalar part of the fermionic supersymmetric transformations
given by
\begin{eqnarray}\label{susytrans3}
S_{\cal AB} &=& \tfrac{1}{2} \e^{K^{\rm v}/2} {V}^\Lambda \Theta_\Lambda^{~~\lambda} P_{\lambda}^x
\varepsilon_\mathcal{AC} (\sigma^x)^{\cal C}_{\cal \ B} \ ,\nonumber\\
W^{i{\cal AB}}
&=& \mathrm{i}\, \e^{K^{\rm v}/2} g^{i\bar \jmath}\,
(\nabla_{\bar \jmath}\bar {V}^\Lambda) \Theta_\Lambda^{~~\lambda} P_{\lambda}^x (\sigma^x)^{\cal A}_{\cal \ C}\, \varepsilon^\mathcal{CB}
\ ,\\
N_\alpha^{\cal A}
&=& 2 \e^{K^{\rm v}/2} \bar {V}^\Lambda \Theta_\Lambda^{~~\lambda} {\cal U}^{\cal A}_{\alpha u} k^u_{\lambda}
\ .\nonumber
\end{eqnarray}
$V^\Lambda$ is a holomorphic symplectic vector defined by $V^\Lambda
\equiv (X^I, \cF_I)$ with K\"ahler-covariant derivative 
$\nabla_i V^\Lambda = \partial_i V^\Lambda + (\partial_i K^{\rm v}) V^\Lambda$.
$\varepsilon_{\cal{AB}}$ is the two-dimensional $\varepsilon$-tensor,
$\varepsilon^{\cal{AB}}$ 
its inverse and $(\sigma^x)^{\cal A}_{\cal \ B}$ are the standard Pauli matrices.
The isometries on ${\M}_{\rm h}$ generated by  $\kk^u_\lambda$ can be 
characterized by a 
triplet of Killing prepotentials (moment maps) $P^x_\lambda$ defined by
\begin{equation}\label{Pdef}
- 2 \kk^u_\lambda\,K_{uv}^x =   \nabla_v P_\lambda^x = \partial_v P_\lambda^x + \epsilon^{xyz} \omega^y_v P_\lambda^z  \ ,
\end{equation}
where $K_{uv}^x $  are the coefficients of the two-forms defined in \eqref{def_Sp(1)_curvature} and $ \omega^y_v$ is the $SU(2)$ connection.
${\mathcal U}^{\mathcal A\alpha}_u $ is the vielbein of ${\M}_{\rm h}$ which can be used to express the metric as
\begin{equation}\label{Udef}
h_{uv} = {\mathcal U}^{\mathcal A\alpha}_u \varepsilon_\mathcal{AB}
\mathcal C_{\alpha \beta} \mathcal U^{\mathcal B\beta}_v  \ ,
\end{equation}
where $\mathcal C_{\alpha \beta}$ is the $Sp(\nh)$ invariant metric.

\subsection{Spontaneous \texorpdfstring{${\cal N} = 2 \rightarrow {\cal N} = 1$}{N = 2 to N = 1} supersymmetry breaking}\label{SSB}

Spontaneous supersymmetry breaking can be analyzed in terms of the
scalar parts of the supersymmetry transformations 
\begin{eqnarray}\label{susytrans2}
\delta_\epsilon \Psi_{\mu {\cal A}} &=& D_\mu \epsilon^*_{\cal A} - S_{\cal AB} \gamma_\mu \epsilon^{\cal B} + \ldots \, ,\nonumber\\
\delta_\epsilon \lambda^{i {\cal A}} &=& W^{i{\cal AB}}\epsilon_{\cal B}+\ldots \, ,\\
\delta_\epsilon \zeta_{\alpha} &=& N_\alpha^{\cal A} \epsilon_{\cal A}+\ldots \, ,\nonumber
\end{eqnarray}
with $S_{\cal AB}, W^{i{\cal AB}}$ and $N_\alpha^{\cal A}$ given in \eqref{susytrans3}
and  
$\epsilon^{\cal A}$ being the $SU(2)$ doublet of 
supersymmetry parameters.
Partial supersymmetry breaking occurs whenever the theory has a background 
where one linear combination
of supersymmetry transformations is non-zero while the second vanishes \cite{Cecotti:1984rk,Cecotti:1984wn}.%
\footnote{This can only be achieved if magnetically charged hypermultiplets are in the spectrum \cite{Ferrara:1995gu,Ferrara:1995xi,Fre:1996js,Bagger:1994vj,Antoniadis:1995vb,LST1}.}
In this case one of the two gravitinos 
gains a mass~$m_{3/2}$ via the super-Higgs mechanism and 
the unbroken supersymmetry implies that this heavy gravitino 
is part of an entire ${\cal N} = 1$ massive spin-3/2 multiplet with spin content $s = (3/2, 1, 1, 1/2)$.
The two massive vectors in this multiplet are the graviphoton together with a 
vector of a vector multiplet
while the necessary Goldstone fields are recruited from two charged 
hypermultiplets \cite{Ferrara:1995gu,Ferrara:1995xi,LST1}. This in turn implies that (at least) two isometries $\kk_{1}$ and $\kk_{2}$ on ${\M}_{\rm h}$ with particular properties have to be gauged.
Concretely, these isometries must be such that for the choice
$P_{1,2}^3 = 0$, the other Killing prepotentials have to satisfy
$P_1^1=-P_2^2$, $P_1^2=P_2^1$ \cite{LST2}.
If additional isometries are gauged, we insist that 
they do not contribute to the
$\cN=2\to\cN=1$ supersymmetry breaking but may  break
$\cN=1$ at some lower scale.

The general solution of partial $\cN=2\to\cN=1$ supersymmetry breaking 
is derived in Refs.~\cite{LST1,LST2} and will not be repeated here in all generality. 
For a given supergravity it constrains the embedding tensor or in other words 
the structure of the gauge charges.  
In this paper we confine our interest to Minkowski backgrounds\footnote{
For examples of $\cN=1$ AdS vacua in this class of theories see e.g.\
\cite{Cassani:2009na}.
} 
where the embedding tensors that partially 
break supersymmetry  can be expressed in the form \cite{LST1}
\begin{equation}\label{solution_embedding_tensor2}
\begin{aligned}
\Theta_I^{\phantom{I}1} = & \Re\big( {\cal F}_{IJ}\,C^J \big) \ , \qquad  \Theta^{I1} =  \Re C^I \ , \\
\Theta_I^{\phantom{I}2} = & \Im\big( {\cal F}_{IJ}\,C^J \big) \ , \qquad  \Theta^{I2} =  \Im  C^I \ ,
\end{aligned}
\end{equation}
with $C^I$ being complex constants satisfying
\begin{align}
\bar C^I (\Im {\cal F})_{IJ} C^J = 0\ .
\label{CtotheI_condition} 
\end{align}
Since the embedding tensor has to be constant, \eqref{solution_embedding_tensor2}
stabilizes  ${\rm rk}({\cal F}_{IJK} C^J)$ of the $\nv$ 
complex coordinates on ${\M}_{\rm v}$ 
by the condition \cite{CLST}
\begin{equation}\label{deformV} 
{\cal F}_{IJ} C^J = \textrm{const.}\qquad\textrm{or}\qquad
{\cal F}_{IJK} C^J \delta X^K = 0 \;.
\end{equation}
This implies that only a  submanifold ${\Mq}_{\rm v}\subset \M_{\rm v}$ descends to the $\cN=1$ theory. 
A similar situation occurs in the hypermultiplet sector as we will see in Section~\ref{Qgeneral}.


Below the scale of supersymmetry breaking one can derive an effective 
low energy
${\cal N} = 1$ theory by
integrating out all massive fields of ${\cal O}(\mino)$.
This includes the massive spin-3/2 multiplet but as we just saw in
\eqref{deformV}
further multiplets can become massive 
and thus
have to be integrated out. The light scalar fields  (denoted by $M^{\hatP}$ in
the following) of 
the resulting effective ${\cal N} = 1$ theory have 
the standard sigma-model couplings\cite{Wess:1992cp}
\begin{eqnarray}\label{N=1Lagrangian}
 {\cal L} ~=~ - \ K_{\hatP  {\bar \hatQ} } D_\mu M^{\hatP} D^{\mu} \bar M^{{\bar \hatQ} } 
- V \ ,
\end{eqnarray}
 where $ K_{\hatP {\bar \hatQ} } = \partial_{\hatP} \bar\partial_{{\bar \hatQ}} K$ 
denotes the K\"ahler metric of the 
scalar field space and $K$ its K\"ahler potential. The  scalar potential is given by
\begin{eqnarray}\label{N=1pot}
  V ~=~ 
  \e^K \big( K^{\hatP {\bar \hatQ}} D_{\hatP} {\cal W} {D_{{\bar \hatQ}} \bar {\cal W}}-3|{\cal W}|^2 \big)
  +\tfrac{1}{2}\,
  (\text{Re}\; f)_{I J} {\cal D}^{ I} {\cal D}^{J}
  \ .
\end{eqnarray}
where ${\cal W}$ is the superpotential and $D_{\hatP} {\cal W} =
\partial_{\hatP} {\cal W} + (\partial_{\hatP} K){\cal W}$ its
K\"ahler-covariant derivative. ${\cal D}^{I}$ are the $\cal D$-terms
for the light ${\cal N} =1$ vector multiplets 
and the holomorphic gauge kinetic function is given by
$f_{I J} = \iu {\cal F}_{I J}$ \cite{LST2}.
%

It was shown in \cite{LST1} that 
integrating out the two massive vector bosons amounts to taking the quotient of ${\M}_{\rm h}$ with respect to the two gauged isometries $\kk_1$ and $\kk_2$ while integrating out the additional massive scalars simply yields a submanifold
of the ${\cal N} =2$ geometry. Therefore 
the scalar field space of the effective ${\cal N} = 1$ theory is given by
\begin{equation}
  \label{N=1product}
 \M^{\cN=1} = {\hat{\M}}_{\rm  h} \times {\hat{\M}}_{\rm  v}\; ,
\end{equation}
where
\begin{equation}\label{quotient}
{\Mq}_{\rm h} \subset {\M}_{\rm    h}/\langle \kk_1, \kk_2 \rangle\ , \qquad {\Mq}_{\rm v} \subset {\M}_{\rm v} \;.
\end{equation}
In Refs.~\cite{LST2,CLST} it was shown that $\M^{\cN=1}$ is indeed a K\"ahler manifold. Its K\"ahler potential is $K = K^{\rm v} + \hat K$ and
we give the explicit form of 
$\hat K$ in the next section for the specific
subclass of K\"ahler manifolds which descend from special quaternionic  K\"ahler manifolds.

If only the two isometries required for partial supersymmetry breaking are gauged, the superpotential and the $\cal D$-terms vanish in a Minkowski background. To get a nontrivial scalar potential, additional isometries have to be gauged at a scale $\tilde m$ below $m_{3/2}$. 
The corresponding superpotential and $\cal D$-terms 
are then given by \cite{LST2} 
\begin{align}
{\cal W} &=  \e ^{- \hat K / 2} V^\Lambda \Theta_\Lambda^{\enspace \lambda} P_\lambda^{\enspace -} \;, \label{superpotential}\\
{\cal D}^{I} &= - \Pi^I_J \Gamma^{J}_K (\Im \cF)^{-1 ~KL} \left(\Theta_{L}^{~~\lambda} - \bar \cF_{LM} \Theta^{M\lambda} \right) P_{\lambda}^3
\;, \label{Dterm}
\end{align}
with $P_\lambda^{\enspace -} \equiv P_\lambda^{\enspace 1} - \iu P_\lambda^{\enspace 2}$ and 
$\Pi^I_J$ and $\Gamma^{J}_K$ are projectors that arise when projecting out the heavy gauge bosons
\begin{equation}\begin{aligned}\label{Pidef}
  \Pi^I_J &= \delta^I_J - 2 \e^{K^{\rm v}} X^I \bar X^K \Im(\cF)_{KJ} \;, \\
  \Gamma_{J}^I &= \delta^I_J - \frac{C^{(P)\,I}  \bar C^{(P)\,K}  \Im(\cF)_{KJ}}{ C^{(P)\,M}  \Im(\cF)_{MN} \bar  C^{(P)\,N}} \;, \qquad \text{with} \ C^{(P)\,I} = \Pi^I_J C^J \;.
\end{aligned}\end{equation}

After this general discussion let us now turn to a specific class of 
${\cal N}=2$ theories
which arise at the tree level of type II string theories compactified on Calabi-Yau threefolds. 

\section{Quotient construction for special K\"ahler manifolds}
\label{sec:specialQuaternionicKaehler}

\subsection{Special quaternionic-K\"ahler manifolds}
\label{sec:specialQuaternionicKaehlersub}

In this paper we focus on the  subclass of ${\cal N} =2$ theories
where ${\M}_{\rm h}$ is restricted to be a special quaternionic-K\"ahler manifold.
Such manifolds are constructed by fibering 
a  specific $(2 \nh +2)$-dimensional $G$-bundle over a $(2 \nh - 2)$-dimensional special-K\"ahler 
submanifold ${\M}_{\rm sk}$.  Let us denote the complex coordinates of
${\M}_{\rm sk}$ by $z^a, a= 1,\ldots,\nh - 1$ and 
 $Z^A = (z^a,1)$, $A= 1,\ldots,\nh$ and the holomorphic prepotential by $\cG(Z)$.
In this notation the K\"ahler potential $K^{\rm h}$ of ${\M}_{\rm sk}$
is given in analogy with 
\eqref{gdef} by
\begin{equation}
K^{\rm h} = -\ln \iu \left(  \bar Z^A \cG_A - Z^A\bar \cG_A \right) 
= -\ln \left(- 2 Z^A N_{AB} \bar Z^B \right)\;,
\label{Kh}
\end{equation}
where we defined
\begin{equation}
  \label{Ndef}
  N_{AB}=\Im {\cal G}_{AB}\ .
\end{equation}
 The (real) coordinates of the fibre are denoted by  $\phi, \ax, \xi^A, \tilde\xi_A$
and the metric is given by \cite{Ferrara:1989ik}
\begin{equation}\begin{aligned}
  \label{qKmetric}
h_{uv}(q)\, \partial_\mu q^u \partial^\mu q^v\  =\ & -(\partial \phi)^2
- e^{4\phi} (\partial \ax +\tilde\xi_A\partial\xi^A - \xi^A\partial\tilde\xi_A )^2
+g_{a\bar b} \partial z^a\partial\bar z^{\bar b}\\
&
+ e^{2\phi} {\Im} {\cal M}^{AB} (\partial\tilde\xi - {\cal M}\partial\xi)_A(\partial\tilde\xi -\bar{\cal M}\partial\xi)_B\ ,
\end{aligned}\end{equation}
where $g_{a\bar b}$ is the metric on ${\M}_{\rm sk}$ and 
\begin{equation}
  \label{Mdef}
  {\cal M}_{AB} = \bar \cG_{AB} +2\iu\ \frac{N_{AC} N_{BD} Z^C Z^D}{N_{DC}  Z^C Z^D}\ .
\end{equation}

The metric \eqref{qKmetric} has
$(2n_{\rm h}+2)$ isometries generated by the Killing vectors
\begin{equation}\label{Killing}
\begin{aligned}
k_{\phi}   ~= &~ \frac{1}{2} \frac{\partial}{\partial \phi} -  \ax \frac{\partial}{\partial \ax} - \frac{1}{2} \xi^A \frac{\partial}{\partial \xi^A} - \frac{1}{2} \tilde \xi_A \frac{\partial}{\partial \tilde \xi_A} \ , \qquad
 k_{\tilde \phi}   ~= ~ - 2 \frac{\partial}{\partial \ax} \ , \\
 k_A  ~= &~ \frac{\partial}{\partial \xi^A} + \tilde \xi_A \frac{\partial}{\partial \ax} \ , \qquad
\tilde k^A  ~= ~ \frac{\partial}{\partial \tilde \xi_A} - \xi^A \frac{\partial}{\partial \ax} \ .
\end{aligned}
\end{equation}
They act transitively on the $G$-fibre coordinates and the subset $\{\Kxi_A, \Ktxi^A,\Kax\} $ spans a Heisenberg algebra which is graded with respect to $k_\phi $. The commutation relations are
\begin{equation}\label{isometries_fibre} 
\begin{aligned}
{} [\Kdil,\Kax]\  =\ & \Kax \ , \qquad 
 [\Kdil,\Kxi_A]\ \ =\  \tfrac{1}{2} \Kxi_A \ , \qquad
 [\Kdil,\Ktxi^A]\  = \ \tfrac{1}{2} \Ktxi^A \ , \qquad 
 [\Kxi_A,\Ktxi^B]\  =  - \delta_A^B \Kax \ ,
\end{aligned}
\end{equation}
with all other commutators vanishing.
Gauged supergravities in this class of theories have been
discussed in \cite{D'Auria:2004tr,D'Auria:2004wd}.

\subsection{Quotient construction for generic base}\label{Qgeneral}

Let us now review the explicit construction of 
$\hat \M_{\rm h} \subseteq {\M}_{\rm    h}/\langle \kk_1, \kk_2 \rangle$
in a Minkowski background with ${\M}_{\rm h}$ being special quaternionic,
 following \cite{LST2,CLST}. Instead of using  $\phi, \ax, \xi^A, \tilde\xi_A$
it is more convenient to use
 another set of coordinates $(z^a, w^0, w_A)$ which are holomorphic on $\M_{\rm h}$ with respect to one of its complex structures, $J^3$.
 $z^a$ are the holomorphic coordinates on ${\M}_{\rm sk}$ as
 introduced 
in Section~\ref{sec:specialQuaternionicKaehlersub}
 while $w^0$ and $w_A$ are given by
\begin{equation} \label{holcoords_Min}\begin{aligned}
 w^0  ~= &~\e^{-2\phi} + \iu (\tilde \phi + \xi^A (\tilde \xi_A - {\cal G}_{AB} \xi^B)) \ , \qquad
 w_A  ~= ~ -\iu (\tilde \xi_A - {\cal G}_{AB} \xi^B) \ .
\end{aligned}\end{equation}
Before we proceed, let us also record  the inverse transformations which are 
given by
\begin{equation}\label{TildeXiOfOmega}
\begin{aligned}
\phi &= - \tfrac{1}{2} \ln \left( \Re w^0  - \Re w_A N^{-1\, AB} \Re w_B \right) \ ,\\
\tilde \xi_A &= -\Re ( {\cal G}_{AB} N^{-1BC} \bar w_C) \ ,  \\
\xi^A &= -\Re (N^{-1AB} \bar w_B) \;. 
\end{aligned}\end{equation}
The K\"ahler potential $\hat K$ of the quotient $\hat \M_{\rm h}$ is derived from
\begin{align}
\hat K = K^h + 2\phi
=  -\ln \iu \left( \bar Z^A \cG_A - Z^A\bar \cG_A \right) - 
\ln \left( \Re w^0  - \Re w_A N^{-1\, AB} \Re w_B \right) \;,
\label{review_hatK_explicit}
\end{align}
where additional constraints have to be imposed.
Moreover, since $\hat \M_{\rm h}$ is a quotient of codimension 2, 
one of the complex coordinates in \eqref{review_hatK_explicit} is redundant.
In the rest of this section we explicitly give the constraints
and their solutions.

The two Killing vectors $\kk_1, \kk_2$ which are relevant for  
the quotient construction are not unique. They are
linear combinations of the Killing vectors \eqref{Killing}
given by
\begin{equation}\label{k1and2}
\begin{aligned}
\kk_1 &= \Re D^A k_A + \Re (D^A {\cal G}_{AB}) \tilde k^B + \Re \left( \iu D^A w_A \right) k_{\tilde \phi}\;,\\
\kk_2 &= \Im D^A k_A + \Im (D^A {\cal G}_{AB}) \tilde k^B + \Im \left( \iu D^A w_A  \right) k_{\tilde \phi}\;,
\end{aligned}
\end{equation}
where the non-uniqueness is parameterized by 
$n_{\rm h}-1$ 
complex constants $D^A$. The condition that $\kk_1$ and $\kk_2$ commute can now be expressed  as
\begin{align}
\bar D^A N_{AB} D^B = 0\ .
\label{DtotheA_condition} 
\end{align}
In addition, all prefactors in \eqref{k1and2} have to be constant 
giving the conditions
\begin{eqnarray}
D^A {\cal G}_{AB} &=& \textrm{const.} 
\;, \label{ZfixingCond}\\
D^A w_A &=& \tilde C 
\label{Ctilde}\; ,
\end{eqnarray}
where  $\tilde C$ are arbitrary complex constants. 
(Note the analogy of \eqref{DtotheA_condition} and \eqref{ZfixingCond}
with \eqref{CtotheI_condition} and \eqref{deformV}.)
The condition \eqref{ZfixingCond}
fixes a subset of scalar fields in the 
special-K\"ahler 
base ${\M}_{\rm sk}$ 
while the condition \eqref{Ctilde}
fixes  one complex scalar field in the fiber. Note that due to \eqref{ZfixingCond} the constraint 
\eqref{DtotheA_condition} does not impose any condition on the scalar
fields, but merely constrains the vector $D$ to lie on the boundary of the domain of the $Z$ coordinates, which is given by $-2 Z^A N_{AB} \bar Z^B =
\e^{-K^{\rm h}} >0$. This also implies that $Z$ and $D$ should not
be proportional to each other
$Z \not\sim D$.

Let us elaborate on the implication of \eqref{ZfixingCond}.
As in  \eqref{deformV} it is equivalent to 
\begin{equation}
{\cal G}_{ABC} D^B \delta Z^C = 0 \ ,
\label{ZfixingCondDeltaZ}
\end{equation}
and thus fixes ${\rm rk}({\cal G}_{ABC} D^B)$ of the $\nh-1$ complex coordinates on ${\M}_{\rm sk}$ \cite{CLST}.
The degree-2 homogeneity of the prepotential implies ${\cal G}_{ABC} Z^C = 0$ and thus
$Z^C$ is a null eigenvector of the $\nh \times \nh$ matrix ${\cal G}_{ABC} D^B$. Therefore it cannot have full rank and we have
\begin{equation}
{\rm rk}({\cal G}_{ABC} D^B) \leq \nh -1\;.
\end{equation}
Denoting by $\nqh$ the number of unfixed complex coordinates on $\M_{\rm sk}$ 
we have
\begin{equation}
\nqh = \nh - 1 - {\rm rk}({\cal G}_{ABC} D^B)\;. 
\end{equation}
Table \ref{tab:baseDimMhHat} displays the corresponding dimensions for some 
typical prepotentials. We see that a generic prepotential of degree three or higher can fix all $\nh - 1$ fields $z^a$,
while for quadratic 
${\cal G}$ \eqref{ZfixingCond} is trivially satisfied  and no $z^a$ are fixed,
since the second derivatives ${\cal G}_{AB}$ are constant.
Intermediate examples where some of the $z^a$ are fixed are given by the cubic $STU$- and quantum-$STU$ prepotentials.
They were discussed in ref.~\cite{CLST} and will be recalled in more detail in Section~\ref{sec:STU}.

\begin{table}[htbp]
\centering
\begin{tabular}{|c|c|c|}
\hline
  ${\cal G}$ & ${\rm rk}({\cal G}_{ABC} D^B)$ &  $\nqh$ \\
  \hline                        
  generic & $\nh - 1$ & 0 \\
  quadratic & 0 & $\nh - 1$ \\
  (quantum) $STU$ & 2 & 1 \\
\hline
\end{tabular}
\caption{Number of base coordinates descending to $\Mq_{\rm h}$ for some typical
prepotentials~${\cal G}$}
\label{tab:baseDimMhHat}
\end{table}

Before we proceed, let us
note that Table \ref{tab:baseDimMhHat} applies to any special K\"ahler manifold that is restricted by the conditions \eqref{DtotheA_condition} and \eqref{ZfixingCond}.
Thus the same argument can be  applied to the vector multiplet sector
as it is determined by the analogous equations \eqref{CtotheI_condition} and \eqref{deformV}.
Let $\nqv \equiv \rm{dim}_{\mathbb{C}}\Mq_{\rm v}$, then Table \ref{tab:baseDimMhHat} can be used  for the vector multiplets with the substitutions
\begin{equation}
{\cal G} \rightarrow  {\cal F}\;, \qquad \quad D^B \rightarrow C^J \;, \qquad \quad  \nh -1 \rightarrow \nv\;, \qquad \quad \nqh \rightarrow \nqv\;. 
\end{equation}

Let us now describe 
the explicit construction of the quotient $\hat \M_{\rm h}$ following
ref.~\cite{CLST}. One first combines the two Killing vectors  
given in \eqref{k1and2} into one holomorphic vector
\begin{equation}
\kkk \equiv \kk_1 - \mathrm{i} \kk_2 = 
-4 \bar D^A \Re w_A \frac{\partial}{\partial w^0} - 2 \bar D^B N_{BA}  \frac{\partial}{\partial w_A}
\end{equation}
where we inserted \eqref{Killing} and used \eqref{holcoords_Min} to change to holomorphic coordinates.
The quotient is taken by identifying 
fiber coordinates on the integral curves generated by $\kkk$
\begin{equation}
\begin{pmatrix} w^0\\ w_A \end{pmatrix} \sim \begin{pmatrix} {w^{0}}'\\ w_A' \end{pmatrix} = \e^{\lambda \kkk} \begin{pmatrix} w^0\\ w_A \end{pmatrix}
 = \begin{pmatrix}w^0 + \lambda \delta_{(1)}^{0} + \lambda^2 \delta_{(2)}^{0} \\ w_A + \lambda \delta_{A} \end{pmatrix}\;,
\label{review_equivalence_Min}
\end{equation}
with $\lambda \in \mathbb{C}$ and
\begin{equation}
\delta_{(1)}^0 = - 4 \bar D^A \Re w_A\;, \qquad \delta_{(2)}^0 = 2\bar D^B N_{BA} \bar D^A\;, \qquad \delta_A = -2 \bar D^B N_{BA}\;.
\label{deltas}
\end{equation}
By inserting \eqref{review_equivalence_Min} into \eqref{review_hatK_explicit} (and using \eqref{DtotheA_condition}), one can check that 
$\kkk$ is indeed a Killing vector as $\hat K$ satisfies
\begin{equation}
\hat K (Z^A, {w^{0}}', w_A') = \hat K (Z^A, {w^{0}}, w_A) \;.
\label{hatKinvariant}
\end{equation}

To remove the coordinate that is redundant on the quotient, 
we use an ansatz similar to Eq.~\eqref{Ctilde}
\begin{equation}\label{Dtilde}
Z^A w_A = \tilde D \;, \quad \tilde D \in \mathbb{C} \;.
\end{equation}
The conditions \eqref{Ctilde} and \eqref{Dtilde} determine two
of the $w_A$ in terms of the other coordinates. Let us denote the two
dependent coordinates by $w_i, w_j,i < j $ while the independent
coordinates are $w_a, a=1,\ldots,\hat i,\ldots,\hat j,\ldots,\nh$
with $\hat i,\hat j$ omitted.
 $w_i, w_j$ are then given by 
\begin{equation}
\begin{aligned}
w_i &= \frac{\left( (Z^j D^a - Z^a D^j) w_a - Z^j \tilde C + D^j \tilde D \right)}{D^j Z^i - D^i Z^j} \ , \label{xCoords}\\
w_j &= -\frac{\left( (Z^i D^a - Z^a D^i) w_a - Z^i \tilde C + D^i \tilde D\right)}{D^j Z^i - D^i Z^j} \ .
\end{aligned}
\end{equation}
Due to \eqref{DtotheA_condition}
$\tilde C$ is constant along the integral curves of  $\kkk$ 
\begin{equation}\begin{aligned}
\tilde C' &= \e^{\lambda \kkk} \tilde C = \tilde C + \lambda D^A \delta_A = \tilde C \;,
\end{aligned}\end{equation}
 so after the elimination of two fiber coordinates, one has to transform the remaining fiber coordinates and $\tilde D$ to stay on the quotient. The K\"ahler potential is of course still invariant under these transformations
\begin{equation}
\hat K (Z^A, {w^{0}}', w_a', \tilde D') = \hat K (Z^A, {w^{0}}, w_a, \tilde D) \;.
\label{hatKinvariant2}
\end{equation}
The ansatz \eqref{Dtilde} was chosen such that the transformation
of $\tilde D$ is always non-zero
\begin{equation}\begin{aligned}
\tilde D' &= \e^{\lambda \kkk} \tilde D = \tilde D + \lambda Z^A \delta_A \;,
\end{aligned}\end{equation}
due to  $Z^A \delta_A \neq 0$ \cite{LST1}.
This guarantees that each equivalence class $[w^0, w_A]$ contains for each $\tilde D \in \mathbb{C}$ exactly one representative,
i.e.\ the quotient is isomorphic to the submanifold obtained by fixing $\tilde D$.
One can now freely choose a $\tilde D$ to pick a representative $(w^0, w_a)$ for the quotient.
In total, 2 of the $\nh + 1$ complex fiber coordinates are fixed in
the ${\cal N} =1$ theory. Including the $\nqh$ remaining base
coordinates, $\Mq_{\rm h}$ thus has complex dimension $\nqh + \nh - 1$.

\section{K\"ahler potential of the quantum \texorpdfstring{$STU$}{STU}-model}
\label{sec:STU}

As an explicit example let us study in detail the quantum $STU$-model
with the prepotential
\begin{equation}\label{STU_prepotential}
{\cal G}=\frac{Z^1 Z^2 Z^3}{Z^4} + \frac{\alpha}{3}\frac{(Z^2)^3}{Z^4}\ 
, \qquad \alpha \in \mathbb{R}  \; .
\end{equation}
It defines  a 16-dimensional  quaternionic K\"ahler manifold
via \eqref{qKmetric} in the original  ${\cal N} = 2$ supergravity
with $\nh=4$ hypermultiplets.
For $\alpha=0$ (called the $STU$-model) the manifold is  the symmetric space
\begin{equation}
\M_{\rm h}=\frac{SO(4,4)}{SO(4) \times SO(4)} \ ,
\end{equation}
with the  six-dimensional special K\"ahler base
\begin{equation}
\M_{\text{sk}}=\left( \frac{SU(1,1)}{U(1)} \right)^3 \;.
\end{equation}
Using \eqref{Kh} and \eqref{Ndef}
one determines the  K\"ahler potential
on the base (before taking the quotient) to be
\begin{equation}
\e^{-K^h} = -2Z^A N_{AB} \bar Z^B = - \tfrac{8}{3}  \Im Z^2 \big(3 \Im Z^1 \Im Z^3 + \alpha (\Im Z^2)^2 \big)\ .
\label{expKhQSTU}
\end{equation}
In order to apply the  quotient construction, let us display
the condition \eqref{ZfixingCond} for the 
prepotential \eqref{STU_prepotential} \cite{CLST}
\begin{equation}\label{qSTUBaseFix}
\begin{aligned}
 (D^4 Z^2 - D^2) (D^4 Z^3 - D^3) &= \mathrm{const.} \ , \\  
(D^4 Z^1 - D^1) (D^4 Z^3 - D^3) + \alpha (D^4 Z^2 - D^2)^2  &= \mathrm{const.} \ , \\
   (D^4 Z^1 - D^1) (D^4 Z^2 - D^2) &= \mathrm{const.}  \ , \\
 (2 D^4 Z^2 - D^2) Z^1 Z^3 - Z^2 (D^1 Z^3 +  Z^1 D^3) + \alpha (Z^2)^2 (\tfrac{2}{3} D^4 Z^2 - D^2) &= \mathrm{const.} \ .
\end{aligned}
\end{equation}
It is easy to see that these equations fix all $Z^A$ unless $Z^2 = \frac{D^2}{D^4}$. With $Z^2$ fixed in this way, the second and fourth equations in \eqref{qSTUBaseFix} both
reduce to
\begin{equation}
(D^4 Z^1 - D^1) (D^4 Z^3 - D^3) = \mathrm{const.} \ ,
\end{equation}
so there are two solutions that leave either $Z^1$ or $Z^3$ free. We pick
\begin{equation}\label{basecon}
Z^1 = \frac{D^1}{D^4}\ ,\qquad  Z^2 = \frac{D^2}{D^4}\ ,\qquad Z^3\ \textrm{arbitrary} \ .
\end{equation}
For simplicity we set $D^4 = 1$ in the following which
can always be achieved by rescaling Eqs.~\eqref{DtotheA_condition}-\eqref{Ctilde}.
The condition \eqref{DtotheA_condition}
that $D^A$ has to be null with respect to $N_{AB}$ then reads
\begin{equation}\label{DtotheA_condition_qSTU}
\Im D^2 \big[ \Im D^1 \Im D^3 + \tfrac{1}{3} \alpha \left(\Im
   D^2\right)^2 \big] = 0\;.
\end{equation}
$\Im D^2\neq 0$ has to hold in order to keep \eqref{expKhQSTU} finite.
This implies that the bracket in \eqref{DtotheA_condition_qSTU} must be zero
and we have 
\begin{equation}
\Im D^1 \neq 0\;, \qquad \Im D^2\neq 0\;, \qquad
\Im D^3 = - \frac{\alpha (\Im D^2)^2}{3 \Im D^1}\;.
\label{qSTUImD1}
\end{equation}
Furthermore, the bracket in \eqref{expKhQSTU} must be non-zero which implies
\begin{equation}
\Im Z^3 \neq - \frac{\alpha (\Im D^2)^2}{3 \Im D^1}\;.
\label{qSTUImZ3}
\end{equation}
In summary we have on the base
\begin{equation}
\begin{gathered}
Z^1 = D^1 \in \mathbb{C}\setminus \mathbb{R} \;,\qquad Z^2 = D^2 \in \mathbb{C}\setminus \mathbb{R}\;,\qquad 
Z^4 = D^4 = 1\;,\\
Z^3 + \iu \frac{\alpha (\Im D^2)^2}{3 \Im D^1} \in \mathbb{C}\setminus \mathbb{R}\;,\qquad D^3 + \iu \frac{\alpha (\Im D^2)^2}{3 \Im D^1} \in \mathbb{R}\;.
\end{gathered}
\label{qZ12fixed}
\end{equation}
Since \eqref{expKhQSTU} must be positive, the choice of $D^A$ determines in which half-plane $Z^3$ has to lie
\begin{equation}
\Im D^1 \Im D^2 \lessgtr 0 \quad \Leftrightarrow \quad \Im Z^3 \gtrless - \frac{\alpha (\Im D^2)^2}{3 \Im D^1}\;.
\label{Z3domains}
\end{equation}

In the fiber two complex coordinates are fixed by \eqref{xCoords}. 
Since $Z$ and $D$ only differ in their third component, the denominator in \eqref{xCoords} is only non-zero if either $i$ or $j$ is 3. One can show \cite{THansen}
that the final result does depend on the specific choice only up to holomorphic coordinate transformations and thus
without loss of generality we can  
set $i = 3, j=4$ in the following to eliminate $w_3$ and $w_4$ in the fiber. 
Furthermore, we also use the fact that we can freely choose a $\tilde D$ and
%
set $\tilde D = \tilde C$ to obtain 
\begin{align}
w_3 = 0, \qquad \quad w_4 = \tilde C - D^1 w_1 - D^2 w_2\;.
\label{w_quotient4}
\end{align}

In order to compute the K\"ahler potential $\hat K$ given in 
\eqref{review_hatK_explicit}
on the quotient 
${\Mq}_{\rm h}$ we first compute 
\begin{eqnarray}\label{NinQSTU}
N^{-1AB} (\alpha)
& = & g(Z)\, h(Z)\,  \Big( 6 (\Im Z^1 \Im Z^3)^2 N^{-1AB} (\alpha=0) \nonumber \\
&& +\; \alpha \Im Z^2 \big( \Re(Z^A \bar Z^B) + 2 \Re Z^A \Re Z^B - 2 \delta^{AB} (\Im Z^A)^2  \\
&& -\; (\delta_1^{A} \delta_3^{B} + \delta_3^{A} \delta_1^{B}) ( 6 \Im Z^1 \Im Z^3 + 2 \alpha (\Im Z^2)^2)  \big) \Big) \;,\nonumber
\end{eqnarray}
where we abbreviated
\begin{equation}\label{ghdef}
g(Z) = \left( 6 \Im Z^1 \Im Z^3 + 2 \alpha (\Im Z^2)^2\right)^{-1} \;, \qquad
h(Z) = \left( \Im Z^1 \Im Z^3 - \alpha (\Im Z^2)^2\right)^{-1} \;,
\end{equation}
and
\begin{equation}
N^{-1AB}(\alpha=0) 
= - \frac{1}{2 \Im Z^1 \Im Z^2 \Im Z^3}
 \begin{pmatrix}
  |Z^1|^2         & \Re ( Z^1 Z^2 ) & \Re ( Z^1 Z^3 ) & \Re ( Z^1 ) \\
  \Re ( Z^1 Z^2 ) & |Z^2|^2         & \Re ( Z^2 Z^3 ) & \Re ( Z^2 ) \\
  \Re ( Z^1 Z^3 ) & \Re ( Z^2 Z^3 ) & |Z^3|^2         & \Re ( Z^3 ) \\
  \Re ( Z^1     ) & \Re ( Z^2     ) & \Re ( Z^3     ) & 1
 \end{pmatrix}.
\label{Nin}
\end{equation}
Inserting \eqref{expKhQSTU}, \eqref{NinQSTU}, \eqref{qZ12fixed} and  \eqref{w_quotient4}
into  \eqref{review_hatK_explicit}, we obtain 
\begin{align}
\e^{-\hat K} = {} & -\tfrac{4}{3} \Im D^2 g(Z^3)^{-1}   \Re w^0  \nonumber \\
& -\; 4 h(Z^3) \Big( \Im D^1 \Im Z^3| \Im D^1 \bar w_1 - \Im D^2 w_2 - \iu \Re \tilde C|^2 \label{qK1}\\
& -\; \tfrac{\alpha}{3} (\Im D^2)^2 \left( 3 (\Re \tilde C + \Im D^a \Im w_a)^2 - (\Im D^1 \Re w_1 - \Im D^2 \Re w_2)^2  \right)  \Big) \;,\nonumber
\end{align}
where
\begin{equation}
g(Z^3) \equiv g(Z^1 = D^1, Z^2 = D^2, Z^3)\;, \qquad h(Z^3) \equiv h(Z^1 = D^1, Z^2 = D^2, Z^3) \;.
\end{equation}
By a set of  field redefinitions -- discussed in Appendix \ref{sec:field_redefinitions} -- $\hat K$ can be brought into the form
\begin{equation}
\e^{-\hat K} = (Z^3 + \bar Z^3) (w^0 + \bar w^0)  - ( w_1 + \bar w_2) ( \bar w_1 + w_2)
+ \frac{\alpha \Big[ (w_1 + \bar w_2) + (\bar w_1 + w_2) \Big]^2}{4\alpha - \frac{3 \Im D^1}{2 (\Im D^2)^2} \left( Z^3 + \bar Z^3 \right)}
\; .
\label{qKhatx1x2}
\end{equation}
For $\alpha=0$ ($STU$-model) we thus obtain
\begin{equation}
\e^{-\hat K} = (Z^3 + \bar Z^3) (w^0 + \bar w^0)  - ( w_1 + \bar w_2) ( \bar w_1 + w_2) \; .
\label{Khatx1x2}
\end{equation}
This is the K\"ahler potential of  the eight-dimensional 
homogeneous K\"ahler manifold \cite{LCLM}
\begin{equation}
\hat \M_{\rm h}=\frac{SO(4,2)}{SO(4) \times SO(2)}\ .
\end{equation} 
%


Let us briefly discuss the fact that the quantum $STU$ K\"ahler potential has a singularity in $Z^3$. In equation \eqref{qK1},
the singularity is present in the form of $h(Z^3)$ and located at $\Im Z^3 = \frac{\alpha (\Im D^2)^2}{\Im D^1}$. It is confined to only one of the two possible domains of $Z^3$ given by \eqref{Z3domains}.

\section{The \texorpdfstring{${\cal N} =1$}{N = 1} scalar potential}\label{WD}
\label{sec:superpot+D-Terms}

\subsection{Special quaternionic manifolds}
To get a non-zero scalar potential,
let us now consider the case where $n > 2$ linearly independent isometries are gauged.
The gauge bosons are recruited among the graviphoton and the vectors of the vector multiplets,
and the number of available gauge bosons limits the number of possible gaugings
to obey 
\begin{equation}\label{gaugingsVsVectorm}
n \leq \nv+1 \; .
\end{equation}
The $n$ commuting Killing vectors are parametrized as in \eqref{k1and2}  by the linear combination
\begin{equation}\label{commutingKilling}
\kk_\lambda = r_\lambda^{\enspace A}\, k_A + s_{\lambda B}\, \tilde k^B + t_\lambda\, k_{\tilde \phi}, \qquad \lambda = 1\ , \ldots, n \; ,
\end{equation}
where $r_\lambda^{\enspace A}, s_{\lambda B},t_\lambda$ are  constant parameters.
$k_{\phi}$ does not appear in \eqref{commutingKilling} 
as 
there are no linear independent commuting Killing vectors involving $k_{\phi}$ (this was shown in \cite{THansen}).
$\kk_{1}$ and $\kk_{2}$ are the two Killing vectors \eqref{k1and2} 
that ensure partial supersymmetry breaking
and thus we have 
\begin{equation}
\begin{aligned}
D^A &= r_{1}^{\enspace A} + \iu r_{2}^{\enspace A} \ , \qquad
D^A {\cal G}_{AB} = s_{1 B}+ \iu s_{2 B} \ .
\label{Ddef}
\end{aligned}
\end{equation}
The additional Killing vectors $\kk_{\lambda>2}$ 
do not participate in the partial supersymmetry breaking, as already discussed
in Section~\ref{SSB}.

$k_A, \tilde k^B, k_{\tilde \phi}$ form a $(2\nh+1)$-dimensional Heisenberg algebra, which has maximal Abelian dimension $(\nh+1)$ \cite{BNT}.
This number is obviously the upper bound for the number of Abelian gaugings
\begin{equation}\label{gaugingsVsHyperm}
n \leq \nh+1 \; .
\end{equation}
Demanding that all $\kk_\lambda$ commute, the commutation relations 
\eqref{isometries_fibre} imply $\tfrac12n(n-1)$ conditions 
on the real parameters  $r_\lambda^{\enspace A}, s_{\lambda B},t_\lambda$
which take a form analogous to the locality constraint \eqref{embeddingTensorLocality} of the embedding tensor 
\begin{equation}\label{killingParameterConds}
r_{[\lambda}^{\enspace A}  s_{\rho] A} = 0 \;.
\end{equation}
%
For 
$\lambda \geq 3$ and $\rho = 1,2$ they can be brought into a more useful form by inserting \eqref{Ddef}
\begin{equation}\label{killingParameterConds12}
(s_{\lambda B} - r_\lambda^{\enspace A}  {\cal G}_{AB} )D^B = 0 \;.
\end{equation}
Similarly, the embedding tensor constraint \eqref{embeddingTensorLocality} with $\lambda \geq 3$ and $\kappa = 1, 2$ reads
\begin{equation}\label{localityCond12}
C^I(\Theta_I^{\enspace \lambda} - {\cal F}_{IJ} \Theta^{J \; \lambda}) = 0 \;,
\end{equation}
after inserting the explicit $\Theta_{\Lambda}^{\phantom{I}1}$ and
$\Theta_{\Lambda}^{\phantom{I}2}$ from
\eqref{solution_embedding_tensor2}.

With coordinates and a basis of Killing vectors on ${\M}_{\rm h}$ at hand, the superpotential \eqref{superpotential} and $\cal D$-terms \eqref{Dterm} can be calculated more explicitly for 
a special quaternionic hypermultiplet sector. 
The only parts of the superpotential and the $\cal D$-terms that depend on the fields of the hypermultiplets are
the Killing prepotentials 
 $P_\lambda^{\enspace -}$ and $P_\lambda^3$, which are calculated in 
Appendix~\ref{sec:the_killing_prepotentials}. 
When inserted, \eqref{superpotential} and \eqref{Dterm} take the form
\begin{align}
{\cal W} &= 2 X^I(\Theta_I^{\enspace \lambda} - {\cal F}_{IJ} \Theta^{J \; \lambda})(s_{\lambda B} - r_\lambda^{\enspace A}  {\cal G}_{AB} )Z^B \;, \label{superpotentialSym}\\
{\cal D}^{I} &= - \e ^{\hat K - K^{\rm h}}  \Pi^I_J \Gamma^{J}_K(\Im {\cal F})^{-1KL}  \left( \Theta_{L}^{~~ \lambda} - \bar{{\cal F}}_{L M} \Theta^{M \lambda}\right)
\Re \left((s_{\lambda B} - r_\lambda^{\enspace A} {\cal G}_{AB})
   N^{-1BC} \bar w_C - t_\lambda\right)
\label{DtermPinserted} .
\end{align}

Since the constraints 
\eqref{killingParameterConds12} and 
\eqref{localityCond12} take a similar form, the superpotential 
\eqref{superpotentialSym} 
is symmetric under the exchange of $(X^I, {\cal F}_I)$ and $(Z^A, {\cal G}_A)$.
This can be made manifest by rewriting \eqref{superpotentialSym} 
in the symplectic form
\begin{align}
{\cal W} &= 2 V^\Lambda \Theta_\Lambda^{\enspace \lambda} s_{\lambda \Sigma}  U^\Sigma, \label{superpotentialSymplectic}
\end{align}
using the symplectic vectors $s_{\lambda \Sigma} \equiv (s_{\lambda A}, -r_\lambda^{\enspace A})$ and $U^\Sigma \equiv (Z^A, {\cal G}_A)$.
If we define the constant matrix $\Theta_{\Lambda\Sigma} \equiv \Theta_\Lambda^{\enspace \lambda} s_{\lambda \Sigma}$ and insert it into  \eqref{superpotentialSymplectic},
we get the superpotential in the same form as given in \cite{LST2}.
The lesson we take away from rederiving the result in the form \eqref{superpotentialSymplectic} is that 
the rank of $\Theta_{\Lambda \Sigma}$ is at most $n - 2$. This can be
seen by recalling that the first two gauged isometries do not
contribute to the superpotential which implies $\lambda = 3, \ldots,
n$ in \eqref{superpotentialSymplectic}. Thus $\Theta_{\Lambda \Sigma}$
is the product of a $(n-2)$-column matrix and a $(n-2)$-row matrix and
its rank is bounded by $n-2$. 

\subsection{\texorpdfstring{$STU$}{STU} and quantum \texorpdfstring{$STU$}{STU} models}
\label{sec:STU_superpotential}

We now evaluate \eqref{superpotentialSym} and \eqref{DtermPinserted}
for the quantum  $STU$ model with an  arbitrary vector multiplet
sector.
Let us start with the superpotential.
It can be simplified by  
using \eqref{killingParameterConds12} and \eqref{qZ12fixed}
which imply
\begin{equation}\label{superpotentialFixZ1Z2}
  (s_{\lambda B} - r_\lambda^{\enspace A} {\cal G}_{AB}) Z^B
=  (s_{\lambda B} - r_\lambda^{\enspace A} {\cal G}_{AB}) (Z^B-D^B) 
=  B^{STU}_\lambda (Z^3 - D^3)\;,
\end{equation}
where we defined the constants
\begin{equation}
B^{STU}_\lambda = s_{\lambda 3} - r_\lambda^{\enspace A} {\cal G}_{A3}\;.
\end{equation}
They are constants since ${\cal G}_{A3}$ does not depend on $Z^3$
\begin{equation}
{\cal G}_{A3} = \left. \partial_A \frac{Z^1 Z^2}{Z^4} \right|_{Z^1=D^1, Z^2=D^2, Z^4=1} = \textrm{const.}\;.
\end{equation}
Inserting into \eqref{superpotentialSymplectic} yields
\begin{align}
{\cal W} &=  2 V^\Lambda \Theta_\Lambda^{\enspace \lambda} B^{STU}_\lambda (Z^3 - D^3)\;. 
\end{align}
Note that the only term which contains a scalar from a hypermultiplet is the
overall factor $(Z^3 - D^3)$. This term can never vanish since
the domains of $Z^3$ and $D^3$ are disjunct due to \eqref{qZ12fixed}.
We will see in the next section that as a consequence, consistency with a Minkowski background requires 
${\cal W}$ to vanish for an appropriate choice 
of prepotentials ${\cal F}$ in the vector multiplet sector.

The $\cal D$-terms for the quantum $STU$ model were calculated in \cite{THansen} and are given by \eqref{DtermPinserted} with the substitution 
\begin{equation}\begin{aligned}
N^{-1BA} \bar w_A =&  - \frac{3 g(Z^3)}{\Im D^2} \Re Z^B \left(  \bar{\tilde C} + \iu \Im D^a \bar w_a \right) \\
& + \frac{h(Z^3)}{2 \Im D^2} \left( \Im Z^B \Im D^a - 2 \delta^{Ba} (\Im D^a)^2\right) \bar w_a\;.
\label{qSTU_Ninw}
\end{aligned}
\end{equation}

\section{Supersymmetric vacua of the \texorpdfstring{${\cal N} = 1$}{N = 1} theory}
\label{sec:generalVacuum}

In this section we study the conditions 
for  supersymmetric vacua of the scalar potential \eqref{N=1pot} which
occur for $\langle D_{U} {\cal W} \rangle = \langle {\cal D}^{I} \rangle = 0$.\footnote{For the non-supersymmetric vacua the analysis is less systematic
and strongly depends on the type of prepotential.}
We do not confine our analysis to the $STU$-models but consider
the prepotentials given in Table~\ref{tab:baseDimMhHat}.
Since throughout this paper we assumed a Minkowskian background, 
consistency  demands that the supersymmetric vacua also have
$\langle {\cal W} \rangle = 0$.

Let us start by showing that this consistency condition is implied by the conditions we stated before. From its
definition  we have $D_{U} {\cal W} = \partial_{U} {\cal W}
+ (\partial_{U} K) {\cal W}$
with $K = \hat K + K^{\rm v}$. 
Since ${\cal W}$ (defined in \eqref{superpotential}) and $K^{\rm v}$ 
(defined in \eqref{gdef}) are independent of $w^0$ and $w_A$, their
partial derivatives with respect to the fiber coordinates vanish and 
thus we have
\begin{equation}
D_{w} {\cal W} =(\partial_{w} \hat K)\, {\cal W}\;.
\end{equation}
$\hat K$ is  given in \eqref{review_hatK_explicit} and we find
\begin{align}
\partial_{w^0} \hat K = - (\Re w^0 - \Re w_A N^{-1AB} \Re w_B)^{-1} = - \e ^{\hat K - K^{\rm h}} \neq 0 \;.
\label{partialOmegaZeroDuK}
\end{align}
As a consequence $\langle D_{w^0} {\cal W} \rangle = 0$ necessarily 
implies $\langle {\cal W} \rangle = 0$.

Let us now turn to the covariant derivatives with respect to the base coordinates $X^{\hat I}$ and $Z^{\hat A}$. 
The index $\hat I$ labels the $\nqv$ ${\cal N} = 1$ scalar fields
descending from the vector multiplets, while $\hat A$ labels the
$\nqh$ 
scalar fields
descending from hypermultiplets. For a (quantum) $STU$
prepotential ${\cal G}$, we only have $Z^3$ on the base, i.e., $\hat A = 3$. A quadratic ${\cal G}$
does not fix any base coordinates and $\hat A = 1, \ldots, \nh - 1$, while a generic ${\cal G}$ fixes all fields. 


The two factors $X^I (\Theta_I^{\enspace \lambda} - {\cal F}_{IJ}
\Theta^{J \; \lambda})$ and $(s_{\lambda B} - r_\lambda^{\enspace A}
{\cal G}_{AB} )Z^B$ that appear in the superpotential
\eqref{superpotentialSym} are both at most linear in the fields $X^{\hat I}$
or $Z^{\hat A}$ for all prepotentials appearing in Table
\ref{tab:baseDimMhHat}.
For the quantum $STU$ prepotential this 
can be explicitly seen from 
\eqref{superpotentialFixZ1Z2}, for quadratic prepotentials ${\cal
  G}_{AB}, {\cal F}_{IJ}$ are constant while generic prepotentials fix
all fields
in the base so that $X^I (\Theta_I^{\enspace \lambda} - {\cal F}_{IJ}
\Theta^{J \; \lambda})$ or $(s_{\lambda B} - r_\lambda^{\enspace A}
{\cal G}_{AB} )Z^B$ are constant altogether.
%
To make this more explicit let us denote the constant parts by 
$A^\lambda$ and $B_\lambda$ respectively so that we have
\begin{equation}\label{cases}
\begin{aligned}
X^I(\Theta_I^{\enspace \lambda} - {\cal F}_{IJ} \Theta^{J \; \lambda})
&\equiv \begin{cases}
(X^3 - C^3) A_{STU}^\lambda\ , \quad & {\cal F}\ \textrm{quantum}\ STU\ , \\
X^I A_{\textrm{quad},I}^\lambda \ , \quad & {\cal F}\ \textrm{quadratic} \ , \\
A_{\textrm{gen}}^\lambda\ , \quad & {\cal F}\ \textrm{generic} \ ,
\end{cases}
\\[1ex]
(s_{\lambda B} - r_\lambda^{\enspace A}  {\cal G}_{AB} ) Z^B
&\equiv \begin{cases}
B^{STU}_\lambda (Z^3 - D^3)\ , \quad & {\cal G}\ \textrm{quantum}\ STU \ , \\
B^{\textrm{quad}}_{\lambda B} Z^B\ , \quad & {\cal G}\ \textrm{quadratic} \ , \\
B^{\textrm{gen}}_\lambda\ , \quad & {\cal G}\ \textrm{generic} \ .
\end{cases}
\end{aligned}
\end{equation}
Altogether we thus have nine possible combinations forming the
superpotential. 
Any of the four combinations of quantum $STU$ and  generic
prepotentials  necessarily have ${\cal W} \equiv 0$.
In these cases the superpotential consists of the 
constant factor $A^\lambda B_\lambda$ possibly multiplied by $(X^3 - C^3)$ or
$(Z^3 - D^3)$ which are non-zero due to \eqref{qZ12fixed}. 
Hence $\langle{\cal W}\rangle = 0$ implies $A^\lambda B_\lambda = 0$ and thus ${\cal W} \equiv 0$. In these cases supersymmetric vacua only exist 
if the superpotential is set to zero by choosing an appropriate embedding tensor and gauged Killing vectors.

The four combinations of a  quadratic prepotential with 
a quantum $STU$ or generic prepotential also only have supersymmetric vacua 
for  ${\cal W} \equiv 0$.
This can be seen from $\langle D_{X^{I}}{\cal W} \rangle = 0$ 
which for ${\cal G}$ generic (and $\langle {\cal W} \rangle=0$) implies
\begin{equation}
\langle \partial_{X^{I}}{\cal W} \rangle = \langle 2
A_{\textrm{quad},I}^\lambda B^{\textrm{gen}} \rangle  = 2
A_{\textrm{quad},I}^\lambda B^{\textrm{gen}} = 0   \;, 
\end{equation}
and thus  ${\cal W} \equiv 0$.
For a quantum $STU$ prepotential ${\cal G}$ one has
\begin{equation}
\langle \partial_{X^{I}}{\cal W} \rangle = \langle 2 A_{\textrm{quad},I}^\lambda B^{STU} (Z^3 - D^3) \rangle = 0 \;. 
\end{equation}
Since again $(Z^3 - D^3)$ cannot vanish $A_{\textrm{quad},I}^\lambda
B^{STU}=0$ and thus ${\cal W} \equiv 0$ has to hold.
We see that in general a quantum $STU$ or generic prepotential 
${\cal  G}$ leads to supersymmetric backgrounds with  
non-trivial superpotentials only  if the ``other'' prepotential ${\cal
  F}$ 
is such that $X^I (\Theta_I^{\enspace \lambda} - {\cal F}_{IJ} \Theta^{J \; \lambda})$ is not just linear in the fields $X^I$.

As the last case we need to discuss both prepotentials being
quadratic.
From \eqref{cases} we learn
\begin{equation}
\langle \partial_{X^{I}}{\cal W} \rangle = \langle 2
A_{\textrm{quad},I}^\lambda B^{\textrm{quad}}_{\lambda A} Z^A \rangle
= 0  \;,\qquad
\langle \partial_{Z^{A}}{\cal W} \rangle = \langle 2X^{I}  A_{\textrm{quad},I}^\lambda B^{\textrm{quad}}_{\lambda A} \rangle =0 \;. \label{dZ_dWZero}
\end{equation}
The conditions \eqref{dZ_dWZero} fix ${\rm rk} (A_{\textrm{quad},I}^\lambda B^{\textrm{quad}}_{\lambda A})$ of the fields $Z^A$ and $X^I$.
$A_{\textrm{quad},I}^\lambda B^{\textrm{quad}}_{\lambda A}$
is the product of a $(\nv+1) \times (n - 2)$ and a $(n-2) \times \nh$ matrix, so its rank is at most ${\rm min}(\nv+1, \nh, n - 2)$.
(The reason why $\lambda$ takes only $n-2$ values is because the first two gauged isometries do not contribute to the superpotential.)
We know from \eqref{killingParameterConds12} and \eqref{localityCond12} that both $A_{\textrm{quad},I}^\lambda$ and $B^{\textrm{quad}}_{\lambda A}$
have a non-trivial null eigenvector
\begin{equation}
C^I A_{\textrm{quad},I}^\lambda = 0\;,\qquad B^{\textrm{quad}}_{\lambda A} D^A = 0 \;,
\end{equation}
which restricts the possible rank of $A_{\textrm{quad},I}^\lambda
B^{\textrm{quad}}_{\lambda A}$ to obey the stronger condition
\begin{equation}
{\rm rk} (A_{\textrm{quad},I}^\lambda B^{\textrm{quad}}_{\lambda A}) \leq {\rm min}(\nv, \nh -1, n - 2 ) = n - 2 \;,
\label{ABmatrixMax}
\end{equation}
where the equality is due to \eqref{gaugingsVsVectorm} and \eqref{gaugingsVsHyperm}.
Now remember that $Z$ can not be proportional to $D$ 
(and analogously $X \not\sim C$). For $A_{\textrm{quad},I}^\lambda B^{\textrm{quad}}_{\lambda A}$
to have two more non-trivial null eigenvectors in addition to $C^I$ and $D^A$, we have to demand
\begin{align}
{\rm rk} (A_{\textrm{quad},I}^\lambda B^{\textrm{quad}}_{\lambda A}) &< \nv \;, \label{quadSuperpotCondition0} \\
{\rm rk} (A_{\textrm{quad},I}^\lambda B^{\textrm{quad}}_{\lambda A}) &< \nh -1 \;. \label{quadSuperpotCondition}
\end{align}
The first condition \eqref{quadSuperpotCondition0} is automatically satisfied due to \eqref{ABmatrixMax} and \eqref{gaugingsVsVectorm}, so \eqref{quadSuperpotCondition} is the consistency condition that has to be imposed 
if there are two quadratic prepotentials.
In this case, there are  $\nh - 1  + \nv - 2 {\rm rk}
(A_{\textrm{quad},I}^\lambda B^{\textrm{quad}}_{\lambda A})$ complex flat
directions among the $X^{I}$ and $Z^{A}$.

Let us summarize the results so far in this section. For
quadratic, quantum $STU$ or generic prepotentials one can have 
non-trivial superpotentials with supersymmetric vacua only when  both 
${\cal G}$ and ${\cal F}$ are quadratic and \eqref{quadSuperpotCondition} holds.

We will now discuss how many fields have to be fixed to satisfy in
addition
the $\cal D$-term condition 
$\langle {\cal D}^{I} \rangle = 0$. 
We first recall that
the projectors $\Pi^I_J$ and $\Gamma^J_K$ defined in \eqref{Pidef}
project out two directions in the field space.
Applied to the ${\cal D}$-terms in \eqref{DtermPinserted} 
one observes that also two of the initially $\nv + 1$  ${\cal D}$-terms
are projected out leaving
at most $\nv - 1$ $\cal D$-terms linearly independent.
In our basis of fermions the $\cal D$-terms are complex
with their phase being a gauge freedom.
So there can be only up to $\nv - 1$ real conditions implied by
$\langle {\cal D}^{I} \rangle = 0$. 
Since the index  $\lambda$ in
\eqref{DtermPinserted}
only takes $n-2$ values,
this number is reduced further
to $n-2$, which is smaller than $\nv - 1$ due to \eqref{gaugingsVsVectorm}.

For $\nh \geq 3$ there are fiber coordinates left after the quotient
construction which can be used to solve the $\cal D$-term condition 
independently
of the prepotentials $\cF$ and $\cG$. 
The $n-2$ real
Killing prepotentials $P_\lambda^3$ (given in \eqref{generalP3}) 
vanish for
\begin{align}\label{Pzero}
\Re \left( (s_{\lambda B} - r_\lambda^{\enspace A} {\cal G}_{AB} ) N^{-1BC} \bar w_C - t_\lambda  \right) = 0\;.
\end{align}
Two of the $w_C$ coordinates are already fixed leaving the $\nh - 2$ 
complex fields $w_a$ to solve \eqref{Pzero}.
Thus, the $n-2$ $P_\lambda^3$ can be set to zero if
$2(\nh - 2) \geq n-2$ which is always satisfied due to \eqref{gaugingsVsHyperm}.
This solution leaves at least $2 \nh - n - 2$ real flat directions among the
$w_a$ on top of the flat directions left in the base after solving the
$F$-term condition. 
In addition $w^0$ is always a complex flat direction.

For $\nh = 2$, $\langle {\cal D}^{ I} \rangle = 0$ 
can only be achieved by fixing
base coordinates $X^I, Z^A$ which may be possible depending on the
form of $\cF$ and $\cG$. However, 
if there is a non-trivial superpotential some of the base coordinates
might already be fixed by the $F$-term condition.
For the quantum $STU$ prepotentials, ${\cal F}_{IJ}$ and ${\cal G}_{AB}$
only depend on  $X^3$ or $Z^3$, respectively 
and these fields can be fixed to set
some of the $\cal D$-terms to zero. For $\nh = 2$, only one additional gauging is allowed, so
all $\cal D$-terms are proportional to just one Killing prepotential
$P_3^3$, which can be set to zero by fixing the real or imaginary part of $Z^3$.

\section{Conclusion}

In this paper, we studied explicit examples
of supergravities which exhibit spontaneous partial
${\cal N}=2\to {\cal N} = 1$ supersymmetry 
breaking in a Minkowskian background.
In the hypermultiplet sector we confined our analysis
to special quaternionic-K\"ahler manifolds which, as we reviewed,
do have isometries that, when appropriately gauged, can induce
the  partial supersymmetry breaking.
We considered  the explicit example
of a special quaternionic-K\"ahler manifold with the (quantum) $STU$ model
as the base. In this case the base is complex three-dimensional
while the fibre is complex five-dimensional.
In the ${\cal N} = 1$ background one of the base and three of the fibre
coordinates span the scalar field space. 
For the (classical) $STU$ model the quaternionic-K\"ahler
manifold is given by $\frac{SO(4,4)}{SO(4) \times SO(4)}$
while 
the  ${\cal N} = 1$ quotient is the K\"ahler manifold
$\frac{SO(4,2)}{SO(4) \times SO(2)}$. For the  quantum $STU$ model
$K$ was explicitly computed but the corresponding K\"ahler 
manifold no longer is a simple symmetric space.

We also considered the situation where additional isometries are gauged
to induce $\cal D$-terms together with a non-trivial superpotential. 
Both depend on the choice of the prepotentials ${\cal F}$ and ${\cal G}$.
We analized the conditions for supersymmetric minima of the scalar potential
for the nine cases where ${\cal F}$ and ${\cal G}$ are quadratic,
quantum $STU$ or generic.
We found that only for ${\cal F}$ and ${\cal G}$ both being quadratic
a non-zero superpotential with supersymmetric minima can exist.
However,  there may exist supersymmetric minima 
for other classes of prepotentials.
The $\cal D$-terms pose no obstruction for supersymmetric vacua if one
considers at least three hypermultiplets.


\section*{Acknowledgments}

We would like to thank Vicente Cortes, Paul Smyth and Hagen Triendl for helpful conversations.
This work was supported by the German Science Foundation (DFG) within the
Collaborative Research Center (SFB) 676 
``Particles, Strings and the Early Universe''.

\vskip2cm

\appendix

\noindent
{\Huge\bf Appendix}

\section{Field redefinitions for the K\"ahler potential}
\label{sec:field_redefinitions}

In this appendix, we display the holomorphic field redefinitions that bring 
the K\"ahler potential \eqref{qK1} into the form \eqref{qKhatx1x2}.
$\tilde C$ in \eqref{qK1} can  be eliminated by the field redefinition 
\begin{equation}\label{fieldRedefinitionsQuantumSTU}
\begin{aligned}
2 \Im D^1 w_1 + \iu \Re \tilde C   \quad &\to \quad w_1\;, \\
- 2 \Im D^2 w_2 - \iu \Re \tilde C   \quad &\to \quad w_2\;, 
\end{aligned}
\end{equation}
resulting in
\begin{equation}
\label{finalQSTUKhat}
\begin{aligned}
\e^{-\hat K} = {} & -g(Z^3)^{-1} \left[ \tfrac{4}{3} \Im D^2 \Re w^0  + \tfrac{1}{6} h(Z^3) (w_1 + \bar w_2) (\bar w_1 + w_2) \right] \\
              & - \tfrac{\alpha}{3} (\Im D^2)^2 h(Z^3) \big[ (w_1 + \bar w_2) - (\bar w_1 + w_2) \big]^2  \;.
\end{aligned}
\end{equation}
This expression can be further simplified by shifting $Z^3$ in such a way that its two domains given by \eqref{Z3domains} are separated by the real axis instead of the line $\Im Z^3 = - \frac{\alpha (\Im D^2)^2}{3 \Im D^1}$
\begin{equation}
Z^3   \quad \to \quad  Z^3 - \iu \frac{\alpha (\Im D^2)^2}{3 \Im D^1}\;.
\end{equation}
In addition we exchange real and imaginary part of $Z^3$ and absorb some constants in $w^0$ by
\begin{equation}\label{fieldRedefinitionsQSTU2}
\begin{aligned}
- \iu Z^3   \quad &\to \quad Z^3\;, \\
- 2 \Im D^1 \Im D^2 w^0   \quad &\to \quad w^0\;.
\end{aligned}
\end{equation}
Inserted into \eqref{ghdef} one finds
\begin{equation}
\begin{aligned}
g(Z^3)   \quad &\to \quad \left( 6 \Im D^1 \Re Z^3 \right)^{-1} \;, \\
h(Z^3)  \quad &\to \quad \left( \Im D^1 \Re Z^3 - \alpha \tfrac{4}{3} (\Im D^2)^2\right)^{-1} \;,
\end{aligned}
\end{equation}
which puts $\hat K$ into the final
form \eqref{qKhatx1x2}.

\section{The Killing prepotentials \texorpdfstring{$P_\lambda^{\enspace -}$ and $P_\lambda^3$}{P- and P3}}
\label{sec:the_killing_prepotentials}

In this appendix we explicitly compute
the Killing prepotentials $P_\lambda^{\enspace -}$ and $P_\lambda^3$
for special quaternionic manifolds. 
In this case the quaternionic vielbein $\mathcal U^{\mathcal A\alpha}$ 
used in \eqref{Udef} reads~\cite{Ferrara:1989ik} 
\begin{equation} \label{quat_vielbein}
\mathcal U^{\mathcal A\alpha}= \mathcal U^{\mathcal A\alpha}_u \diff q^u =\tfrac{1}{\sqrt{2}}
\left(\begin{aligned}
 \bar{u} && \bar{e} && -v && -E \\
\bar{v} && \bar{E} && u && e
\end{aligned}\right) \ ,
 \end{equation}
with the one-forms
\begin{equation} \label{one-forms_quat}
 \begin{aligned}
  u ~= &~ \iu \e^{K^{\rm h}/2+\phi}Z^A(\diff \tilde\xi_A - \mathcal M_{AB} \diff \xi^B) \ , \\
  v ~= &~ \tfrac{1}{2} \e^{2\phi}\big[ \diff \e^{-2\phi}-\iu (\diff \ax +\tilde\xi_A \diff \xi^A-\xi^A \diff \tilde \xi_A  ) \big] \ , \\
  E^{\,\underline{b}} ~= &~ -\tfrac{\iu}{2} \e^{\phi-K^{\rm h}/2} {\Proj}_A^{\phantom{A}\underline{b}} N^{-1AB}(\diff \tilde\xi_B - \mathcal M_{BC} \diff \xi^C) \ , \\
  e^{\,\underline{b}} ~= &~ {\Proj}_A^{\phantom{A}\underline{b}} \diff Z^A \ . \\
 \end{aligned}
\end{equation}
${\Proj}_A^{\phantom{A}\underline{b}} =(-e_a^{\phantom{a}\underline{b}}Z^a, e_a^{\phantom{a}\underline{b}})$ is defined using the vielbein $e_a^{\phantom{a}\underline{b}}$ on $\M_{\rm  sk}$ and
$\mathcal M_{AB}$ is defined in \eqref{Mdef}.
The $SU(2)$ connections $\omega^x$ are given by
\begin{equation}\label{quat_connection}
\begin{aligned}
\omega^1 ~= &~  \iu (\bar u- u)  \ , \qquad
\omega^2 ~= ~ u + \bar u \ , \\
\omega^3 ~= &~ \tfrac{\iu}{2} (v-\bar v) - \iu \e^{K^{\rm h}} \left(Z^A N_{AB} \diff \bar Z^B - \bar Z^A N_{AB} \diff Z^B \right) \ .
 \end{aligned}
\end{equation}

The Killing prepotentials $P^x_\lambda$ of a special quaternionic-K\"ahler manifold take the simple form \cite{Michelson:1996pn}
\begin{equation} \label{prepotential_no_compensator}
 P^x_\lambda = \omega^x_u k_\lambda^u \ .
\end{equation}
Inserting the Killing vectors \eqref{commutingKilling}
in terms of the basis vectors \eqref{Killing} we find
\begin{align}
P_\lambda^{\enspace -} \equiv P_\lambda^{\enspace 1} - \iu P_\lambda^{\enspace 2}
&= (\omega_u^{1}- \iu \omega_u^{2})\, \kk_\lambda^u \nonumber \\
&= -2 \iu u_u \left(r_\lambda^{\enspace A} k_A^u + s_{\lambda A} \tilde k^{Au} + t_\lambda k_{\tilde \phi}^u \right)  \label{generalPminus} \\
&= 2 \e ^{\hat K / 2} (s_{\lambda B} - r_\lambda^{\enspace A} {\cal G}_{AB}) Z^B \;, \nonumber
\end{align}
where we used the identity $Z^A {\cal M}_{AB} = Z^A {\cal G}_{AB}$ that follows from the definition of ${\cal M}_{AB}$.
For $P_\lambda^3$ we obtain analogously
\begin{align}
P_\lambda^3 = \omega^3_u \kk_\lambda^u = \e ^{2 \phi} \left( r_\lambda^{\enspace A} \tilde \xi_A -  s_{\lambda A} \xi^A - t_\lambda  \right) \;.
\end{align}
Using  \eqref{TildeXiOfOmega}
one can  switch to holomorphic coordinates
\begin{align}
P_\lambda^3 = \e ^{2 \phi} \Re \left( (s_{\lambda B} - r_\lambda^{\enspace A} {\cal G}_{AB} ) N^{-1BC} \bar w_C - t_\lambda  \right) \label{generalP3}\;.
\end{align}

\providecommand{\href}[2]{#2}\begingroup\raggedright

\end{document}